\documentclass[onecolumn]{article}
\usepackage{graphicx}
\usepackage[utf8]{inputenc}
\usepackage{amsmath}
\usepackage{physics}
\usepackage{mathtools}
\usepackage{amssymb}
\usepackage{slashed}
\usepackage{tikz-feynman}
\usepackage{amsthm}
\usepackage{txfonts}
\usepackage{geometry}
\begin{document}

\title{A Higgs Doublet + Higgs Singlet Scheme with Negative Kinetic Term for Neutrino Mass Generation}
\author{Lu Jianlong\\ \emph{\small{Department of Physics, National University of Singapore}}}
\date{}
\maketitle

\begin{abstract}
In this paper, we propose a scheme for neutrino mass generation. A Higgs singlet is introduced in the Standard Model of particle physics (SM). All of this new singlet and the charged and neutral Higgs fields in the original Higgs doublet are assumed to have nonzero vacuum expectation values (VEVs). The normal hierarchy of neutrino masses is evidently preferred in our scheme. The neutrino masses are explicitly calculated.

\end{abstract}

\section{INTRODUCTION}
At the present stage, the SM is widely accepted as the unified scheme to understand phenomena of the strong, electromagnetic and weak interactions, which is based on the gauge symmetry group ${\rm SU}(3)_{c}\times {\rm SU}(2)_{L}\times {\rm U}(1)_{Y}$. All elementary fermions and gauge bosons in the SM would be massless if the whole gauge symmetry is unbroken. By incorporating a Higgs doublet $\begin{pmatrix}\phi^{+}\\ \phi^{0}\end{pmatrix}$ with a nonzero VEV in its neutral component $\phi^{0}$, the electroweak gauge symmetry ${\rm SU}(2)_{L}\times {\rm U}(1)_{Y}$ is spontaneously broken into ${\rm U}(1)_{Q}$. All weak gauge bosons ($W^{+}$, $W^{-}$ and $Z$) and all elementary fermions, except neutrinos, obtain nonzero masses via the so-called Higgs mechanism. To be specific, the masses of weak gauge bosons come from the covariant derivative acting on the Higgs doublet, and the masses of massive elementary fermions originate from the Yukawa couplings with the Higgs doublet. There exist no right-handed (left-handed) components of neutrino (antineutrino) fields in the SM Lagrangian, thus neutrinos (antineutrinos) are assumed to be massless. This picture, although phenomenologically successful, has been strongly questioned due to the observed flavor-changing phenomena of neutrinos. One natural solution to this problem, an analog of the mixing picture in the quark sector, is to assign nonzero masses to neutrinos such that the neutrino mass eigenstates mismatch its flavor eigenstates.\cite{Maki} However, it immediately brought us a new problem: \emph{Where do the neutrino masses come from?} There have been many different proposals, such as the seesaw mechanism. In the next section we propose a new model of neutrino mass generation, in which there is not need of any unknown heavy neutrinos as compared to seesaw mechanism. The normal mass hierarchy is predicted in our 
model by referring to the experimental values of charged lepton masses. The neutrino masses are explicitly calculated, with reasonable orders of magnitude.

\section{THE MODEL}
In our model, there are intrinsically two differences compared with the unextended SM. First, the charged Higgs field in the Higgs doublet is supposed to have a nonzero VEV. Second, there exists one additional Higgs singlet $S$, which takes part in the game by the following covariant derivative,
\begin{equation}
   D_{\mu}S=\Big(\partial_{\mu}-\sum_{a=1}^{3}ig_{2}T^{a}W_{\mu}^{a}-ig_{1}\frac{Y}{2}B_{\mu}\Big)S,
\end{equation}
where $T^{a}$ is the matrix representation of the $a$-th generator of ${\rm SU}(2)$, and $Y=1$. Recall that the generators of ${\rm SU}(2)$ must satisfy $[T^{a},T^{b}]=i\varepsilon^{abc}T^{c}$. In $2$-dimensional case about the Higgs doublet, as discussed in \cite{Lu}, we have $T^{a}=\frac{\sigma_{a}}{2}$ because
\begin{equation}
    \big[\frac{\sigma_{a}}{2},\frac{\sigma_{b}}{2}\big]=\frac{1}{4}[\sigma_{a},\sigma_{b}]=i\varepsilon^{abc}\frac{\sigma_{c}}{2}.
\end{equation}
To apply the covariant derivative to the Higgs singlet $S$, we need $1$-dimensional representation for the generators of ${\rm SU}(2)$. A nontrivial choice is based on quaternions. We let $T^{1}=i\frac{\textbf{i}}{2}$, $T^{2}=i\frac{\textbf{j}}{2}$ and $T^{3}=i\frac{\textbf{k}}{2}$, emphasizing the fundamental quaternion units in boldface. The commutation relations $[T^{a},T^{b}]=i\varepsilon^{abc}T^{c}$ are obviously satisfied. The kinetic term of $S$ is assumed to differ from the ordinary kinetic terms of other fields in the SM by an additional minus sign, called negative kinetic term, i.e.,
\begin{equation}
    \mathcal{L}^{S}_{{\rm kinetic}}= -(D^{\mu}S)^{\dagger} D_{\mu}S.
\end{equation}
The notion negative kinetic term is also discussed in literatures of cosmology (e.g. \cite{carroll}). By perturbing $S$ around one of its VEVs, denoted by $N$, 
\begin{equation}
    S=N+q(x),
\end{equation}  
$\mathcal{L}^{S}_{{\rm kinetic}}$ can be explicitly written as 
\begin{equation}
\begin{gathered}
    \mathcal{L}^{S}_{{\rm kinetic}}= -(D^{\mu}S)^{\dagger} D_{\mu}S\\
   =-\Big(\partial^{\mu}S+\sum_{a=1}^{3}ig_{2}T^{a*}SW^{\mu}_{a}+ig_{1}\frac{S}{2}B^{\mu}\Big) \Big(\partial_{\mu}S-\sum_{b=1}^{3}ig_{2}T^{b}SW_{\mu}^{b}-ig_{1}\frac{S}{2}B_{\mu}\Big)\\
   =-\Big(\partial^{\mu}q+\sum_{a=1}^{3}ig_{2}T^{a*}(N+q)W^{\mu}_{a}+ig_{1}\frac{N+q}{2}B^{\mu}\Big) \Big(\partial_{\mu}q-\sum_{b=1}^{3}ig_{2}T^{b}(N+q)W_{\mu}^{b}-ig_{1}\frac{N+q}{2}B_{\mu}\Big)\\
   = -\partial^{\mu}q\partial_{\mu}q-\sum_{a=1}^{3}g^{2}_{2}\frac{(N+q)^{2}}{4}W^{\mu}_{a}W_{\mu}^{a}-g^{2}_{1}\frac{(N+q)^{2}}{4}B^{\mu}B_{\mu}\\
   = -\partial^{\mu}q\partial_{\mu}q-g^{2}_{2}\frac{(N+q)^{2}}{4}(W^{\mu}_{1}+i W^{\mu}_{2})(W_{\mu}^{1}-i W_{\mu}^{2})-\frac{(N+q)^{2}}{8}(g_{2}W_{3}^{\mu}+g_{1}B^{\mu})(g_{2}W^{3}_{\mu}+g_{1}B_{\mu})\\-       \frac{(N+q)^{2}}{8}(g_{2}W_{3}^{\mu}-g_{1}B^{\mu})(g_{2}W^{3}_{\mu}-g_{1}B_{\mu}).
\end{gathered}
\end{equation}
Recall that the four gauge bosons of electroweak interaction are defined as follows:
\begin{equation}
   W^{\pm}_{\mu}=\frac{1}{\sqrt{2}}(W_{\mu}^{1}\mp iW_{\mu}^{2}),\ \ \ \ Z_{\mu}=\frac{g_{2}W_{\mu}^{3}-g_{1}B_{\mu}}{\sqrt{g_{2}^{2}+g_{1}^{2}}},\ \ \ \ A_{\mu}=\frac{g_{2}W_{\mu}^{3}+g_{1}B_{\mu}}{\sqrt{g_{2}^{2}+g_{1}^{2}}}.
\end{equation}
The terms responsible for masses of electroweak gauge bosons in the above $\mathcal{L}^{S}_{{\rm kinetic}}$ are 
\begin{equation}
   -g^{2}_{2}\frac{N^{2}}{4}(W^{\mu}_{1}+i W^{\mu}_{2})(W_{\mu}^{1}-i W_{\mu}^{2})-\frac{N^{2}}{8}(g_{2}W_{3}^{\mu}+g_{1}B^{\mu})(g_{2}W^{3}_{\mu}+g_{1}B_{\mu})-       \frac{N^{2}}{8}(g_{2}W_{3}^{\mu}-g_{1}B^{\mu})(g_{2}W^{3}_{\mu}-g_{1}B_{\mu}),
\end{equation}
which can be simplified as
\begin{equation}
    -g^{2}_{2}\frac{N^{2}}{2}W^{-\mu}W_{\mu}^{+}-\frac{N^{2}(g_{2}^{2}+g_{1}^{2})}{8}A^{\mu}A_{\mu}-       \frac{N^{2}(g_{2}^{2}+g_{1}^{2})}{8}Z^{\mu}Z_{\mu}.
\end{equation}
In \cite{Lu}, we have derived the electroweak gauge boson mass terms coming from the nonzero VEVs of neutral and charged Higgs fields,
\begin{equation}
   \frac{g_{2}^{2}(\eta^{2}+|\eta^{+}|^{2})}{8} (W_{1}^{\mu}+iW_{2}^{\mu}) (W_{\mu}^{1}-iW^{2}_{\mu})+ \frac{|\eta^{+}|^{2}}{8} (g_{2}W^{\mu}_{3}+g_{1}B^{\mu}) (g_{2}W_{\mu}^{3}+g_{1}B_{\mu})+ \frac{\eta^{2}}{8}(g_{2}W^{\mu}_{3}-g_{1}B^{\mu} )(g_{2}W_{\mu}^{3}-g_{1}B_{\mu}),
\end{equation}
i.e.,
\begin{equation}
   \frac{g_{2}^{2}(\eta^{2}+|\eta^{+}|^{2})}{4} W^{-\mu}W_{\mu}^{+} + \frac{|\eta^{+}|^{2}(g_{2}^{2}+g_{1}^{2})}{8} A^{\mu}A_{\mu}+ \frac{\eta^{2}(g_{2}^{2}+g_{1}^{2})}{8}Z^{\mu}Z_{\mu},
\end{equation}
where $\eta$ and $\eta^{+}$ are respectively the VEVs of neutral and charged Higgs fields. The combination of contributions from Higgs doublet $\Phi$ and Higgs singlet $S$ is then 
\begin{equation}
   \frac{g_{2}^{2}(\eta^{2}+|\eta^{+}|^{2}-2N^{2})}{4} W^{-\mu}W_{\mu}^{+} + \frac{(|\eta^{+}|^{2}-N^{2})(g_{2}^{2}+g_{1}^{2})}{8} A^{\mu}A_{\mu}+ \frac{(\eta^{2}-N^{2})(g_{2}^{2}+g_{1}^{2})}{8}Z^{\mu}Z_{\mu}.
\end{equation}
If $N=|\eta^{+}|$, the photon remains massless, while the masses of $W^{\pm}$ and $Z$ bosons are modified accordingly. We have shown in \cite{Lu} that a Higgs doublet with nonzero VEVs in its charged and neutral components is not enough to generate neutrino masses satisfyingly, because it will lead to a photon mass much larger than the current experimental upper bound. That is why we need $S$ to provide offset to the photon mass. In the unextended SM, the left-handed components of charged lepton and neutrino in the same generation together form a doublet, leaving the right-handed component of charged lepton as a singlet. Their Yukawa couplings with the Higgs doublet produce nonzero masses of charged leptons. In our model, we can show that the new Higgs singlet $S$ does not contribute to the lepton masses via Yukawa coupling. Consider the mass dimensions of fermion fields ($\frac{3}{2}$) and scalar fields ($1$), the renormalizable gauge invariant Yukawa coupling terms among $S$ and lepton fields we can write down are 
\begin{equation}
    \mathcal{L}^{{\rm singlet}}_{{\rm Yukawa}}=-\sum_{l=e,\mu,\tau}\Big[y_{1}\bar{l}_{R}Sl_{R} +y_{2}\begin{pmatrix}\bar{\nu}_{lL} & \bar{l}_{L}\end{pmatrix}S \begin{pmatrix}\nu_{lL} \\ l_{L}\end{pmatrix}+{\rm h.c.}\Big],
\end{equation}
in which $y_{1}$ and $y_{2}$ are Yukawa coupling constants. The left-handed component $\psi_{L}$ and right-handed component $\psi_{R}$ of a Dirac field $\psi$ are defined as 
\begin{equation}
    \psi_{L}=\frac{1-\gamma^{5}}{2}\psi,\ \ \ \ \psi_{R}= \frac{1+\gamma^{5}}{2}\psi.
\end{equation}
Thus the Dirac adjoints of $l_{L}$, $l_{R}$ and $\nu_{lL}$ are respectively 
\begin{equation}
   \bar{l}_{L}=(l_{L})^{\dagger}\gamma^{0}=\big(\frac{1-\gamma^{5}}{2}l\big)^{\dagger}\gamma^{0}=l^{\dagger} \frac{1-\gamma^{5}}{2}\gamma^{0},
\end{equation}
\begin{equation}
   \bar{l}_{R}= (l_{R})^{\dagger}\gamma^{0}=\big(\frac{1+\gamma^{5}}{2}l\big)^{\dagger}\gamma^{0}=l^{\dagger} \frac{1+\gamma^{5}}{2}\gamma^{0},
\end{equation}
\begin{equation}
   \bar{\nu}_ {lL}= (\nu_{lL})^{\dagger}\gamma^{0}=\big(\frac{1-\gamma^{5}}{2}\nu_{l}\big)^{\dagger}\gamma^{0}=\nu_{l}^{\dagger} \frac{1-\gamma^{5}}{2}\gamma^{0}.
\end{equation}
They immediately imply a vanishing $\mathcal{L}^{{\rm singlet}}_{{\rm Yukawa}}$ because of 
\begin{equation}
    y_{1}\bar{l}_{R}Sl_{R}=y_{1} l^{\dagger} \frac{1+\gamma^{5}}{2}\gamma^{0}S \frac{1+\gamma^{5}}{2}l= y_{1} l^{\dagger}\gamma_{0}S \frac{1-\gamma^{5}}{2} \frac{1+\gamma^{5}}{2}l=0,
\end{equation}
\begin{equation}
    y_{2}\begin{pmatrix}\bar{\nu}_{lL} & \bar{l}_{L}\end{pmatrix}S \begin{pmatrix}\nu_{lL} \\ l_{L}\end{pmatrix}=y_{2} \bar{\nu}_{lL}S \nu_{lL}+y_{2} \bar{l}_{L}S l_{L}= y_{2} \nu_{l}^{\dagger} \frac{1-\gamma^{5}}{2}\gamma^{0}S \frac{1-\gamma^{5}}{2}\nu_{l}+y_{2} l^{\dagger} \frac{1-\gamma^{5}}{2}\gamma^{0}S \frac{1-\gamma^{5}}{2}l =0.
\end{equation}
Note that we have made use of $\{\gamma^{5},\gamma^{0}\}=0$ and $(\gamma^{5})^{-1}=\gamma^{5}$ in the above derivation.\\
For the original Higgs doublet, the Yukawa coupling terms with three generations of charged leptons and neutrinos are 
\begin{equation}
   \mathcal{L}_{{\rm Yukawa}}^{L}=-\sum_{l=e,\mu,\tau}\Big[y_{l}\bar{l}_{R}\Phi^{\dagger}\begin{pmatrix}\nu_{lL}\\ l_{L}\end{pmatrix}+{\rm h.c.}\Big],
\end{equation}
where $y_{l}$ is the Yukawa coupling constant for the $l$ generation of leptons. As mentioned previously, we assume that both neutral Higgs field and charged Higgs field, which together form the Higgs doublet, have nonzero VEVs. Thus we make the following substitution,
\begin{equation}
   \Phi=\frac{1}{\sqrt{2}}\begin{pmatrix}\eta^{+}\\ \eta+h\end{pmatrix}.
\end{equation}
Note that we have implicitly made use of the ${\rm SU}(2)$ gauge symmetry of the theory to eliminate three unphysical degrees of freedom in the above Higgs doublet. In other words, the unitary gauge has been chosen. Direct substitution gives us
\begin{equation}
\begin{gathered}
   \mathcal{L}_{{\rm Yukawa}}^{L}=-\sum_{l=e,\mu,\tau}\Big[y_{l}\bar{l}_{R}\Phi^{\dagger}\begin{pmatrix}\nu_{lL}\\ l_{L}\end{pmatrix}+{\rm h.c.}\Big]\\
    = -\sum_{l=e,\mu,\tau}\Big[y_{l}\bar{l}_{R}\Phi^{\dagger}\begin{pmatrix}\nu_{lL}\\ l_{L}\end{pmatrix}+ y_{l} \begin{pmatrix}\bar{\nu}_{lL} & \bar{l}_{L}\end{pmatrix}  \Phi l_{R}       \Big]\\
    = -\sum_{l=e,\mu,\tau}\Big[y_{l}\bar{l}_{R} \frac{1}{\sqrt{2}}\begin{pmatrix}\eta^{+*} & \eta+h\end{pmatrix}\begin{pmatrix}\nu_{lL}\\ l_{L}\end{pmatrix}+ y_{l} \begin{pmatrix}\bar{\nu}_{lL} & \bar{l}_{L}\end{pmatrix}  \frac{1}{\sqrt{2}}\begin{pmatrix}\eta^{+}\\ \eta+h\end{pmatrix} l_{R}       \Big]\\
    =- \sum_{l=e,\mu,\tau} \Big[ \frac{1}{\sqrt{2}} y_{l}\eta^{+*}\bar{l}_{R} \nu_{lL}+ \frac{1}{\sqrt{2}} y_{l}\eta\bar{l}_{R} l_{L}+ \frac{1}{\sqrt{2}} y_{l}h\bar{l}_{R} l_{L}+\frac{1}{\sqrt{2}}y_{l}\eta^{+} \bar{\nu}_{lL} l_{R}  + \frac{1}{\sqrt{2}}y_{l}\eta \bar{l}_{L} l_{R}  + \frac{1}{\sqrt{2}}y_{l}h \bar{l}_{L} l_{R}\Big].
\end{gathered}
\end{equation}
According to the definitions of left-handed (right-handed) component and Dirac adjoint of Dirac fields, $\mathcal{L}_{{\rm Yukawa}}^{L}$ can be expressed in terms of only $l$ and $\nu_{lL}$, i.e.,
\begin{equation}
\begin{gathered}
   \mathcal{L}_{{\rm Yukawa}}^{L} =- \sum_{l=e,\mu,\tau} \Big[ \frac{1}{\sqrt{2}} y_{l}\eta^{+*}\bar{l}_{R} \nu_{lL}+ \frac{1}{\sqrt{2}} y_{l}\eta\bar{l}_{R} l_{L}+ \frac{1}{\sqrt{2}} y_{l}h\bar{l}_{R} l_{L}+\frac{1}{\sqrt{2}}y_{l}\eta^{+} \bar{\nu}_{lL} l_{R}  + \frac{1}{\sqrt{2}}y_{l}\eta \bar{l}_{L} l_{R}  + \frac{1}{\sqrt{2}}y_{l}h \bar{l}_{L} l_{R}\Big]\\
    =- \sum_{l=e,\mu,\tau} \Big[ \frac{1}{\sqrt{2}} y_{l}\eta^{+*}\bar{l} \nu_{lL}+ \frac{1}{\sqrt{2}} y_{l}\eta\bar{l} l_{L}+ \frac{1}{\sqrt{2}} y_{l}h\bar{l} l_{L}+\frac{1}{\sqrt{2}}y_{l}\eta^{+} \bar{\nu}_{lL} l+ \frac{1}{\sqrt{2}}y_{l}\eta \bar{l}l_{R}  + \frac{1}{\sqrt{2}}y_{l}h \bar{l} l_{R}\Big]\\
    = - \sum_{l=e,\mu,\tau} \Big[ \frac{1}{\sqrt{2}} y_{l}\eta^{+*}\bar{l} \nu_{lL}+\frac{1}{\sqrt{2}}y_{l}\eta^{+} \bar{\nu}_{lL} l+ \frac{1}{\sqrt{2}}y_{l}\eta \bar{l}l  + \frac{1}{\sqrt{2}}y_{l}h \bar{l} l\Big].
\end{gathered}
\end{equation}
Note that we have implicitly used $\{\gamma^{5},\gamma^{0}\}=0$ and $\frac{1\pm\gamma^{5}}{2} \frac{1\pm\gamma^{5}}{2}= \frac{1\pm\gamma^{5}}{2}$. It is clear that there is no mass terms for the left-handed neutrino fields $\nu_{lL}$. However, the existence of non-diagonal terms $\frac{1}{\sqrt{2}} y_{l}\eta^{+*}\bar{l} \nu_{lL}$ and $\frac{1}{\sqrt{2}}y_{l}\eta^{+} \bar{\nu}_{lL} l$ immediately motivates us to diagonalize the matrix
\begin{equation}
\begin{pmatrix}
   0 & \frac{1}{\sqrt{2}}y_{l}\eta^{+}\\
   \frac{1}{\sqrt{2}} y_{l}\eta^{+*} &  \frac{1}{\sqrt{2}}y_{l}\eta
\end{pmatrix}.
\end{equation}
This can be achieved by 
\begin{equation}
   \begin{pmatrix}
   0 & \frac{1}{\sqrt{2}}y_{l}\eta^{+}\\
   \frac{1}{\sqrt{2}} y_{l}\eta^{+*} &  \frac{1}{\sqrt{2}}y_{l}\eta
\end{pmatrix}=\begin{pmatrix} -\frac{\eta+\sqrt{\eta^{2}+4|\eta^{+}|^{2}}   }{2 \eta^{+*} }  &   \frac{-\eta+\sqrt{\eta^{2}+4|\eta^{+}|^{2}}   }{2 \eta^{+*} }   \\ 1 & 1\end{pmatrix} \begin{pmatrix} \frac{y_{l}\eta-y_{l}\sqrt{\eta^{2}+4|\eta^{+}|^{2}}   }{2\sqrt{2} }    & 0\\ 0 & \frac{y_{l}\eta+y_{l}\sqrt{\eta^{2}+4|\eta^{+}|^{2}}   }{2\sqrt{2} }\end{pmatrix} \begin{pmatrix} -\frac{\eta^{+*}}{\sqrt{\eta^{2}+4|\eta^{+}|^{2}}} &   \frac{1}{2}-\frac{\eta}{\sqrt{\eta^{2}+4|\eta^{+}|^{2}}}   \\  \frac{\eta^{+*}}{\sqrt{\eta^{2}+4|\eta^{+}|^{2}}}   &   \frac{1}{2}+\frac{\eta}{\sqrt{\eta^{2}+4|\eta^{+}|^{2}}}     \end{pmatrix}.
\end{equation}
Obviously, this is not the only way to diagonalize the mass matrix, since we can introduce new degrees of freedom by inserting two identity matrices:
\begin{equation}
\begin{gathered}
   \begin{pmatrix}
   0 & \frac{1}{\sqrt{2}}y_{l}\eta^{+}\\
   \frac{1}{\sqrt{2}} y_{l}\eta^{+*} &  \frac{1}{\sqrt{2}}y_{l}\eta
\end{pmatrix}=\begin{pmatrix} -\frac{\eta+\sqrt{\eta^{2}+4|\eta^{+}|^{2}}   }{2 \eta^{+*} }  &   \frac{-\eta+\sqrt{\eta^{2}+4|\eta^{+}|^{2}}   }{2 \eta^{+*} }   \\ 1 & 1\end{pmatrix}\begin{pmatrix}1 & 0\\ 0 & 1\end{pmatrix} \begin{pmatrix} \frac{y_{l}\eta-y_{l}\sqrt{\eta^{2}+4|\eta^{+}|^{2}}   }{2\sqrt{2} }    & 0\\ 0 & \frac{y_{l}\eta+y_{l}\sqrt{\eta^{2}+4|\eta^{+}|^{2}}   }{2\sqrt{2} }\end{pmatrix}\\ \times   \begin{pmatrix}1 & 0\\ 0 & 1\end{pmatrix} \begin{pmatrix} -\frac{\eta^{+*}}{\sqrt{\eta^{2}+4|\eta^{+}|^{2}}} &   \frac{1}{2}-\frac{\eta}{2\sqrt{\eta^{2}+4|\eta^{+}|^{2}}}   \\  \frac{\eta^{+*}}{\sqrt{\eta^{2}+4|\eta^{+}|^{2}}}   &   \frac{1}{2}+\frac{\eta}{2\sqrt{\eta^{2}+4|\eta^{+}|^{2}}}     \end{pmatrix}\\
    = \begin{pmatrix} -\frac{\eta+\sqrt{\eta^{2}+4|\eta^{+}|^{2}}   }{2 \eta^{+*} }  &   \frac{-\eta+\sqrt{\eta^{2}+4|\eta^{+}|^{2}}   }{2 \eta^{+*} }   \\ 1 & 1\end{pmatrix} \begin{pmatrix}\frac{1}{a} & 0\\ 0 & \frac{1}{b}\end{pmatrix} \begin{pmatrix}a & 0\\ 0 & b\end{pmatrix} \begin{pmatrix} \frac{y_{l}\eta-y_{l}\sqrt{\eta^{2}+4|\eta^{+}|^{2}}   }{2\sqrt{2} }    & 0\\ 0 & \frac{y_{l}\eta+y_{l}\sqrt{\eta^{2}+4|\eta^{+}|^{2}}   }{2\sqrt{2} }\end{pmatrix}\\ \times  \begin{pmatrix}\frac{1}{a} & 0\\ 0 & \frac{1}{b}\end{pmatrix} \begin{pmatrix}a & 0\\ 0 & b\end{pmatrix} \begin{pmatrix} -\frac{\eta^{+*}}{\sqrt{\eta^{2}+4|\eta^{+}|^{2}}} &   \frac{1}{2}-\frac{\eta}{2\sqrt{\eta^{2}+4|\eta^{+}|^{2}}}   \\  \frac{\eta^{+*}}{\sqrt{\eta^{2}+4|\eta^{+}|^{2}}}   &   \frac{1}{2}+\frac{\eta}{2\sqrt{\eta^{2}+4|\eta^{+}|^{2}}}     \end{pmatrix}\\
    = \begin{pmatrix} -\frac{\eta+\sqrt{\eta^{2}+4|\eta^{+}|^{2}}   }{2 \eta^{+*} }  &   \frac{-\eta+\sqrt{\eta^{2}+4|\eta^{+}|^{2}}   }{2 \eta^{+*} }   \\ 1 & 1\end{pmatrix} \begin{pmatrix}\frac{1}{a} & 0\\ 0 & \frac{1}{b}\end{pmatrix} \begin{pmatrix} \frac{y_{l}\eta-y_{l}\sqrt{\eta^{2}+4|\eta^{+}|^{2}}   }{2\sqrt{2} }    & 0\\ 0 & \frac{y_{l}\eta+y_{l}\sqrt{\eta^{2}+4|\eta^{+}|^{2}}   }{2\sqrt{2} }\end{pmatrix} \begin{pmatrix}a & 0\\ 0 & b\end{pmatrix}   \\ \times  \begin{pmatrix}\frac{1}{a} & 0\\ 0 & \frac{1}{b}\end{pmatrix} \begin{pmatrix}a & 0\\ 0 & b\end{pmatrix} \begin{pmatrix} -\frac{\eta^{+*}}{\sqrt{\eta^{2}+4|\eta^{+}|^{2}}} &   \frac{1}{2}-\frac{\eta}{2\sqrt{\eta^{2}+4|\eta^{+}|^{2}}}   \\  \frac{\eta^{+*}}{\sqrt{\eta^{2}+4|\eta^{+}|^{2}}}   &   \frac{1}{2}+\frac{\eta}{2\sqrt{\eta^{2}+4|\eta^{+}|^{2}}}     \end{pmatrix}\\
    = \begin{pmatrix} -\frac{\eta+\sqrt{\eta^{2}+4|\eta^{+}|^{2}}   }{2 \eta^{+*} }  &   \frac{-\eta+\sqrt{\eta^{2}+4|\eta^{+}|^{2}}   }{2 \eta^{+*} }   \\ 1 & 1\end{pmatrix} \begin{pmatrix}\frac{1}{a} & 0\\ 0 & \frac{1}{b}\end{pmatrix} \begin{pmatrix} \frac{y_{l}\eta-y_{l}\sqrt{\eta^{2}+4|\eta^{+}|^{2}}   }{2\sqrt{2} }    & 0\\ 0 & \frac{y_{l}\eta+y_{l}\sqrt{\eta^{2}+4|\eta^{+}|^{2}}   }{2\sqrt{2} }\end{pmatrix}\begin{pmatrix}a & 0\\ 0 & b\end{pmatrix} \begin{pmatrix} -\frac{\eta^{+*}}{\sqrt{\eta^{2}+4|\eta^{+}|^{2}}} &   \frac{1}{2}-\frac{\eta}{2\sqrt{\eta^{2}+4|\eta^{+}|^{2}}}   \\  \frac{\eta^{+*}}{\sqrt{\eta^{2}+4|\eta^{+}|^{2}}}   &   \frac{1}{2}+\frac{\eta}{2\sqrt{\eta^{2}+4|\eta^{+}|^{2}}}     \end{pmatrix}
\end{gathered}
\end{equation}
where $a$ and $b$ are nonzero complex numbers. Not all degrees of freedom are legitimate here. In order to have appropriate mass terms, we must have 
\begin{equation}
   -\frac{1}{a} \frac{\eta+\sqrt{\eta^{2}+4|\eta^{+}|^{2}}   }{2 \eta^{+*} }\bar{\nu}_{lL}+\frac{1}{a}\bar{l}=\Big(-a\frac{\eta^{+*}}{\sqrt{\eta^{2}+4|\eta^{+}|^{2}}} \nu_{lL}+  a\big( \frac{1}{2}-\frac{\eta}{2\sqrt{\eta^{2}+4|\eta^{+}|^{2}}}\big)l      \Big)^{\dagger}\gamma^{0},
\end{equation}
\begin{equation}
   \frac{1}{b} \frac{-\eta+\sqrt{\eta^{2}+4|\eta^{+}|^{2}}   }{2 \eta^{+*} } \bar{\nu}_{lL}+\frac{1}{b}\bar{l}=\Big( b\frac{\eta^{+*}}{\sqrt{\eta^{2}+4|\eta^{+}|^{2}}}\nu_{lL}   +   b\big(\frac{1}{2}+\frac{\eta}{2\sqrt{\eta^{2}+4|\eta^{+}|^{2}}}\big)l\Big)^{\dagger}\gamma^{0},
\end{equation}
i.e.,
\begin{equation}
   -\frac{1}{a} \frac{\eta+\sqrt{\eta^{2}+4|\eta^{+}|^{2}}   }{2 \eta^{+*} }\bar{\nu}_{lL}+\frac{1}{a}\bar{l}=-a^{*}\frac{\eta^{+}}{\sqrt{\eta^{2}+4|\eta^{+}|^{2}}} \bar{\nu}_{lL}+  a^{*}\big( \frac{1}{2}-\frac{\eta}{2\sqrt{\eta^{2}+4|\eta^{+}|^{2}}}\big)\bar{l},
\end{equation}
\begin{equation}
   \frac{1}{b} \frac{-\eta+\sqrt{\eta^{2}+4|\eta^{+}|^{2}}   }{2 \eta^{+*} } \bar{\nu}_{lL}+\frac{1}{b}\bar{l}= b^{*}\frac{\eta^{+}}{\sqrt{\eta^{2}+4|\eta^{+}|^{2}}}\bar{\nu}_{lL}   +   b^{*}\big(\frac{1}{2}+\frac{\eta}{2\sqrt{\eta^{2}+4|\eta^{+}|^{2}}}\big)\bar{l}.
\end{equation}
Equating corresponding coefficients gives
\begin{equation}
     |a|^{2}=2+\frac{1}{2}\frac{\eta^{2}}{|\eta^{+}|^{2}}+\frac{1}{2}\sqrt{\frac{\eta^{4}}{|\eta^{+}|^{4}}+4 \frac{\eta^{2}}{|\eta^{+}|^{2}}},
\end{equation}
\begin{equation}
     |b|^{2}= 2+\frac{1}{2}\frac{\eta^{2}}{|\eta^{+}|^{2}}-\frac{1}{2}\sqrt{\frac{\eta^{4}}{|\eta^{+}|^{4}}+4 \frac{\eta^{2}}{|\eta^{+}|^{2}}}.
\end{equation}
The remaining unrestricted degrees of freedom are the phases of $a$ and $b$. Now we define two new fields $\nu_{lM}$ and $l'$ by mixing $\nu_{lL}$ and $l$,
\begin{equation}
    \begin{pmatrix} \nu_{lM} \\ l'\end{pmatrix} = \begin{pmatrix} -\frac{a\eta^{+*}}{\sqrt{\eta^{2}+4|\eta^{+}|^{2}}} &   a\big(\frac{1}{2}-\frac{\eta}{2\sqrt{\eta^{2}+4|\eta^{+}|^{2}}}\big)   \\  \frac{b\eta^{+*}}{\sqrt{\eta^{2}+4|\eta^{+}|^{2}}}   &   b\big(\frac{1}{2}+\frac{\eta}{2\sqrt{\eta^{2}+4|\eta^{+}|^{2}}}\big)     \end{pmatrix} \begin{pmatrix} \nu_{lL} \\ l\end{pmatrix}.
\end{equation}
We call this mixing matrix $\Omega$, with $\Omega_{ij}$ as its $ij$-entry. The subscript $L$ of the neutrino fields has dropped out because $\frac{1-\gamma^{5}}{2}\nu_{lM}\neq \nu_{lM}$. With these two new fields, the previous Yukawa coupling terms become
\begin{equation}
    \mathcal{L}_{{\rm Yukawa}}^{L}= - \sum_{l=e,\mu,\tau} \Big[\begin{pmatrix}\bar{\nu}_{lM} & \bar{l}'\end{pmatrix}  \begin{pmatrix} \frac{y_{l}\eta-y_{l}\sqrt{\eta^{2}+4|\eta^{+}|^{2}}   }{2\sqrt{2} }    & 0\\ 0 & \frac{y_{l}\eta+y_{l}\sqrt{\eta^{2}+4|\eta^{+}|^{2}}   }{2\sqrt{2} }\end{pmatrix} \begin{pmatrix} \nu_{lM} \\ l'\end{pmatrix} + \frac{1}{\sqrt{2}}y_{l}h \bar{l} l\Big].
\end{equation}
It is easy to notice that $\frac{y_{l}\eta-y_{l}\sqrt{\eta^{2}+4|\eta^{+}|^{2}}   }{2\sqrt{2} }$ is negative for any nonzero $\eta^{+}$ assuming $y_{l}$ is positive, which cannot be a physical mass. We can solve this problem with the help of the relations $\{\gamma^{0},\gamma^{5}\}=0$ and $(\gamma^{5})^{-1}=\gamma^{5}=(\gamma^{5})^{\dagger}$. First, we take away one minus sign from the first eigenvalue by
\begin{equation}
\begin{gathered}
     \begin{pmatrix}\bar{\nu}_{lM} & \bar{l}'\end{pmatrix}  \begin{pmatrix} \frac{y_{l}\eta-y_{l}\sqrt{\eta^{2}+4|\eta^{+}|^{2}}   }{2\sqrt{2} }    & 0\\ 0 & \frac{y_{l}\eta+y_{l}\sqrt{\eta^{2}+4|\eta^{+}|^{2}}   }{2\sqrt{2} }\end{pmatrix} \begin{pmatrix} \nu_{lM} \\ l'\end{pmatrix}= \begin{pmatrix}\bar{\nu}_{lM} & \bar{l}'\end{pmatrix}\begin{pmatrix}-c^{-1} & 0\\ 0 & 1\end{pmatrix}  \begin{pmatrix} \frac{y_{l}\sqrt{\eta^{2}+4|\eta^{+}|^{2}}-y_{l}\eta   }{2\sqrt{2} }    & 0\\ 0 & \frac{y_{l}\eta+y_{l}\sqrt{\eta^{2}+4|\eta^{+}|^{2}}   }{2\sqrt{2} }\end{pmatrix}  \begin{pmatrix}c & 0\\ 0 & 1\end{pmatrix}   \begin{pmatrix} \nu_{lM} \\ l'\end{pmatrix}.
\end{gathered}
\end{equation}
Then we have 
\begin{equation}
     -\bar{\nu}_{lM}c^{-1}=\big( c\nu_{lM}\big)^{\dagger}\gamma^{0},
\end{equation}
i.e.,
\begin{equation}
     -\nu^{\dagger}_{lM}\gamma^{0}c^{-1}=\nu^{\dagger}_{lM}c^{\dagger}\gamma^{0}.
\end{equation}
This can be satisfied if $c=i\gamma^{5}$. Hence, we define another new field $\nu'_{lM}=i\gamma^{5}\nu_{lM}$. The Yukawa coupling terms eventually become
\begin{equation}
    \mathcal{L}_{{\rm Yukawa}}^{L}= - \sum_{l=e,\mu,\tau} \Big[\begin{pmatrix}\bar{\nu}'_{lM} & \bar{l}'\end{pmatrix}  \begin{pmatrix} \frac{y_{l}\sqrt{\eta^{2}+4|\eta^{+}|^{2}}-y_{l}\eta   }{2\sqrt{2} }    & 0\\ 0 & \frac{y_{l}\eta+y_{l}\sqrt{\eta^{2}+4|\eta^{+}|^{2}}   }{2\sqrt{2} }\end{pmatrix} \begin{pmatrix} \nu'_{lM} \\ l'\end{pmatrix} + \frac{1}{\sqrt{2}}y_{l}h \bar{l} l\Big].
\end{equation}
We can see that in this scheme the states of charged leptons interacting with $h$ are different from those charged lepton mass eigenstates. As discussed previously, the new Higgs singlet does not contribute to the lepton masses. Hence, the mass eigenvalues of charged lepton are $\frac{y_{l}\eta+y_{l}\sqrt{\eta^{2}+4|\eta^{+}|^{2}}   }{2\sqrt{2} }$ with $l=e,\mu,\tau$, denoted by $m_{l}$, which are slightly modified from the original results in the unextended SM, $\frac{y_{l}\eta}{\sqrt{2}}$. The neutrino masses predicted in this scheme are $\frac{y_{l}\sqrt{\eta^{2}+4|\eta^{+}|^{2}}-y_{l}\eta   }{2\sqrt{2} }$. One crucial implication is that the ratio among neutrino masses is the same as the ratio among charged lepton masses,
\begin{equation}
\begin{gathered}
     \frac{y_{e}\eta+y_{e}\sqrt{\eta^{2}+4|\eta^{+}|^{2}}   }{2\sqrt{2} }: \frac{y_{\mu}\eta+y_{\mu}\sqrt{\eta^{2}+4|\eta^{+}|^{2}}   }{2\sqrt{2} }: \frac{y_{\tau}\eta+y_{\tau}\sqrt{\eta^{2}+4|\eta^{+}|^{2}}   }{2\sqrt{2} }\\ = \frac{y_{e}\sqrt{\eta^{2}+4|\eta^{+}|^{2}}-y_{e}\eta   }{2\sqrt{2} }: \frac{y_{\mu}\sqrt{\eta^{2}+4|\eta^{+}|^{2}}-y_{\mu}\eta   }{2\sqrt{2} }: \frac{y_{\tau}\sqrt{\eta^{2}+4|\eta^{+}|^{2}}-y_{\tau}\eta   }{2\sqrt{2} }\\ =y_{e}:y_{\mu}:y_{\tau}.
\end{gathered}
\end{equation}
The neutrino mass hierarchy has not been determined yet due to the unknown sign of $\Delta m_{32}^{2}$. \cite{XZZ} There are two possible neutrino mass hierarchy, the normal hierarchy ($m_{1}<m_{2}<m_{3}$) and the inverted hierarchy ($m_{3}< m_{1}<m_{2}$). The known pattern of lepton masses, listed below \cite{pm1}, clearly indicates the preference for the normal hierarchy in our model, which means $m_{1}= \frac{y_{e}\sqrt{\eta^{2}+4|\eta^{+}|^{2}}-y_{e}\eta   }{2\sqrt{2} }$, $m_{2}= \frac{y_{\mu}\sqrt{\eta^{2}+4|\eta^{+}|^{2}}-y_{\mu}\eta   }{2\sqrt{2} }$ and $m_{3}= \frac{y_{\tau}\sqrt{\eta^{2}+4|\eta^{+}|^{2}}-y_{\tau}\eta   }{2\sqrt{2} }$.
\begin{equation}
\begin{gathered}
    m_{e}\approx 0.5109989461 \ {\rm MeV},\\
    m_{\mu}\approx 105.6583745 \ {\rm MeV},\\
    m_{\tau}\approx 1776.86  \ {\rm MeV}.\\
\end{gathered}
\end{equation}
The experimental values of neutrino mass-squared differences are also needed for calculating neutrino masses, which are quoted as follows \cite{pm2}:
\begin{equation}
\begin{gathered}
    \Delta m_{21}^{2}\approx 7.53\times 10^{-5} \ {\rm eV}^{2},\\
    \Delta m_{32}^{2}\approx 2.44\times 10^{-3} \ {\rm eV}^{2}.
\end{gathered}
\end{equation}
Here we have chosen the values of $\Delta m_{32}^{2}$ obtained under the assumption of normal mass hierarchy. The calculation of neutrino masses can be based on either $(m_{e},m_{\mu},m_{\tau},\Delta m_{21}^{2})$ or $(m_{e},m_{\mu},m_{\tau},\Delta m_{32}^{2})$. The results are:
\begin{equation}
    m_{1}\approx 4.20\times 10^{-5} \ {\rm eV},\ \ m_{2}\approx 8.68\times 10^{-3} \ {\rm eV},\ \ m_{3}\approx 1.46\times 10^{-1} \ {\rm eV}\ \ \ \ \ ({\rm based\ on}\ \Delta m_{21}^{2})
\end{equation}
\begin{equation}
    m_{1}\approx 1.42\times 10^{-5} \ {\rm eV},\ \ m_{2}\approx 2.94\times 10^{-3} \ {\rm eV},\ \ m_{3}\approx 4.95\times 10^{-2} \ {\rm eV}\ \ \ \ \ ({\rm based\ on}\ \Delta m_{32}^{2})
\end{equation}
Remarkably, the predicted values of neutrino masses from two sets of parameters differ from each other only by a small factor less than $3$. Referring to the experimental upper bounds of the sum of neutrino masses listed in $\cite{pm1}$, we can see that the above prediction is indeed reasonable.\\
To dock with the theory of neutrino mixing and oscillation, we should have 
\begin{equation}
    \begin{pmatrix}\nu_{eL} \\ \nu_{\mu L}\\ \nu_{\tau L}\end{pmatrix}=\begin{pmatrix} U_{e1} & U_{e2} & U_{e3}\\ U_{\mu 1} & U_{\mu 2} & U_{\mu 3}\\ U_{\tau 1} & U_{\tau 2} & U_{\tau 3}\end{pmatrix} \begin{pmatrix}\nu_{1} \\ \nu_{2}\\ \nu_{3}\end{pmatrix}
\end{equation}
with $(\nu_{1},\nu_{2},\nu_{3})=(\nu'_{eM},\nu'_{\mu M},\nu'_{\tau M})$ assuming normal hierarchy of neutrino masses. Then the original lepton fields ($\nu_{eL}$, $\nu_{\mu L}$, $\nu_{\tau L}$ and $e$, $\mu$, $\tau$) are related by 
\begin{equation}
    \nu_{lL}=  U_{l1}i\gamma^{5}(\Omega_{11}\nu_{eL}+\Omega_{12}e)+ U_{l2}i\gamma^{5}(\Omega_{11}\nu_{\mu L}+\Omega_{12}\mu)+ U_{l3}i\gamma^{5}(\Omega_{11}\nu_{\tau L}+\Omega_{12}\tau).
\end{equation}
The factor $\gamma^{5}$ can be absorbed by utilizing $\gamma^{5}\psi_{L}=-\psi_{L}$ and $\gamma^{5}\psi_{R}=\psi_{R}$,
\begin{equation}
    \nu_{lL}=  U_{l1}i\Big(-\Omega_{11}\nu_{eL}+\Omega_{12}(e_{R}-e_{L})\Big)+ U_{l2}i\Big(-\Omega_{11}\nu_{\mu L}+\Omega_{12}(\mu_{R}-\mu_{L})\Big)+ U_{l3}i\Big(-\Omega_{11}\nu_{\tau L}+\Omega_{12}(\tau_{R}-\tau_{L})\Big).
\end{equation}
Evidently, those nine fields ($\nu_{e}$, $\nu_{\mu}$, $\nu_{\tau}$, $e_{L}$, $e_{R}$, $\mu_{L}$, $\mu_{R}$, $\tau_{L}$, $\tau_{R}$) are not totally independent in our model.

\section{CONCLUSION}
We have constructed a scheme for neutrino mass generation based on a Higgs doublet and a Higgs singlet together with three nonzero VEVs, which predicts the neutrino masses at reasonable orders of magnitude. The masses of weak interaction gauge bosons acquire small modifications, while photon remains massless. The smallness of neutrino masses originates from the smallness of VEV of the charged Higgs field. Normal hierarchy of neutrino masses is revealed by the known pattern of charged lepton masses.


\end{document}